# $V_S$ structure of the crust and upper mantle in the Balkan Peninsula region.

*Reneta B. Raykova*[1] *and Giuliano F. Panza*[2, 3, 4, 5]

[1]Sofia University "St. Kl. Ohridski", Faculty of Physics, Department of Meteorology and Geophysics, Blvd. J. Bauchier 5, Sofia; rraykova@phys.uni-sofia.bg
[2]University of Trieste, Department of Mathematics and Geosciences, via E. Weiss 4, Trieste, Italy; panza@units.it
[3]The Abdus Salam International Centre for Theoretical Physics, SAND Group, Trieste – Italy
[4]International Seismic Safety Organization (ISSO) - www.issoquake.org
[5]Institute of Geophysics, China Earthquake Administration, Beijing

## Introduction

Estimations of Earth's structure have been performed in the last several decades on a variety of scales. The most common application of surface-wave dispersion tomography deals with the improvement of global and regional velocity models of the Earth's crust and upper mantle. The inversion of broadband regional surface-wave velocity maps provides more detailed and realistic picture of the Earth's lithosphere with respect to studies that use globally propagating surface waves or observation of teleseismic body waves alone.

This study defines a 3D shear-wave velocity model to the depth of about 350 km in the region of Balkan Peninsula by application of several methods and tools: collection of surface-wave dispersion measurements; tomography on a grid sized 1°x1°; non-linear inversion of dispersion relations; local smoothing optimization; and juxtaposition of representative cellular models. This study upgrades, refines and extends the results obtained for the region by Raykova and Panza (2006), Raykova and Nikolova (2007), Panza et al. (2007) and Brandmayr et al. (2010).

## Surface-wave dispersion data

The main part of the collected data is composed by group-velocity measurements made by Pontevivo and Panza (2001, 2006), Karagniani et al. (2002), Raykova et al. (2004), Raykova and Nikolova (2007) and El Gabry et al. (2013). The frequency-time analysis (FTAN, Levshin et al., 1989) was used to extract the fundamental mode of Rayleigh waves and to calculate relevant dispersion curve from each seismogram. Additionally, published group and phase velocity measurements were included to increase the density and penetration depth of the data. The resulting data set for Rayleigh wave dispersions consists of more than 1000 group velocity measurements that span over the period range from about 5 s to 80 s and more than 150 phase velocity measurements in the period range from 10 s to 150 s.

## 2D tomography

The two-dimensional tomography based on the Backus–Gilbert method was used to determine the local values of the group and/or phase velocities for a set of periods, mapping the horizontal (at a specific period) and vertical (at a specific grid knot) variations in the Earth's structure. The choice of the set of periods is based on the vertical resolving power of the available data, as determined by the partial derivatives of dispersion values (group and phase velocities) with respect to the structural parameters (Panza, 1981, Urban et al., 1993). The lateral resolution of the tomographic maps is defined as the average size of the equivalent smoothing area and its elongation (Yanovskaya, 1997) and hence it is not necessary to perform check board or similar tests (Foulger et al., 2013).

## Local dispersion curves

Local values of group and phase velocities were assembled for cells sized 1° x 1°. Group-velocity data from the global tomography study of Ritzwoller and Levshin (1998) were used to extend the period range of group velocities from 80 s to 150 s. The local dispersion curves constructed in such a way span over a varying period range according to the availability of the data. Each local dispersion value is qualified by an error that is combination of measurement error and tomographic resolution.

## Non-linear inversion

The non-linear "hedgehog" inversion of cellular dispersion curves was applied to obtain the shear-wave velocity models for cells sized 1°×1° in the Balkan Peninsula region. The period range of dispersion data allows to retrieve velocity structure reliably in the depth range from 3 km – 10 km to about 350 km. The structural model of each cell was modeled as a stack of 19 homogeneous elastic isotropic layers and some of the layers were replaced by a finite number of numerical parameters. A priori independent information about the crustal properties of each defined cell was used in parameterization of layers to improve the resolving power of the tomographic data (Pontevivo and Panza, 2006). Uppermost five layers have properties fixed (specific for each cell). The underlying five layers have variable S- and P-wave velocity, VS and VP, thicknesses and if appropriate – density. VP, VS, and density are fixed in the eleventh layer, while the thickness varies so that the whole stack of eleven layers has a total thickness of 350 km. The bottom layers, below 350 km, consist of Poissonian material (Du et al,. 1998), are common to all cells and are kept fixed during inversion. The thickness of the uppermost crustal layers and thickness step in the parameterized crustal layers were rounded to 0.5 km or 1 km (the only exception is the water layer rounded to 0.1 km, a precision consistent with bathymetry resolution).

The applied non-linear inversion method is a trial-and-error optimized Monte Carlo search and the group and phase velocities of the Rayleigh waves (fundamental mode) were computed for each tested structural model. The model is accepted if the difference between the computed and measured values at each period is less than the single error at the relevant period and if the root mean square (r.m.s.) values for the whole group and phase velocity curves are less than the given limits.

The non-linear inversion is multi-valued and a set of models was obtained as the solution for each cell. The number of the solutions is controlled by average r.m.s. value of the cellular dispersion curves for phase and group velocities and in general not exceeded 15 models per cell.

## Local smoothness optimization (LSO) algorithm

A representative model, with its uncertainties, for each cell is required in order to construct a 3D model of the studied area and to define the geological meaning of the resulting structures. An optimized smoothing method (Boyadzhiev et al., 2008), which follows the Occam's razor principle, was used to define the representative model for each cell with a formalized criterion. This method decreases the introduction of artificial vertical discontinuities in VS models between neighboring cells and keeps the 3D structure as homogeneous as possible, minimizing the lateral velocity gradient and the dependence of the final model from the predefined grid. The models are represented as velocity vectors with equal size for all sets of cellular solutions. The divergence between two models (vectors) is defined as the standard Euclidean norm between two vectors. Each velocity value was weighted with the ratio between layer thickness and the total thickness of the structure.

LSO looks for a solution of the inverse problem, searching cell by cell, considering only the neighbors of the selected cell and fixing the solution as the one, which minimizes the norm between such neighbors. The algorithm follows the principle of minimal divergence in the models. The search starts from the cell with minimal average dispersion of the cellular models, where cellular solutions are the densest in the parameter space and therefore the potential systematic bias introduced by the choice is minimized. The LSO selects the cellular model, minimizing the norm between the neighbor cells (one side in common), as the representative solution of the processed cell. Once a solution is chosen in the running cell, the search continues by applying the procedure to the neighboring cell, not yet processed, with the smallest average dispersion of the cellular models. The direction of "maximum stability" is followed in the progressive choice of the representative solution of the cells until the whole investigation domain is explored.

Since the non-linear inversion and its smoothing optimization guarantee only the mathematical validity of the solution of the inverse problem, the optimization procedure may be repeated whenever necessary, including additional geophysical constrains, such as Moho boundary depth, seismic energy distribution versus depth, presence of magmatism, heat flow, etc.

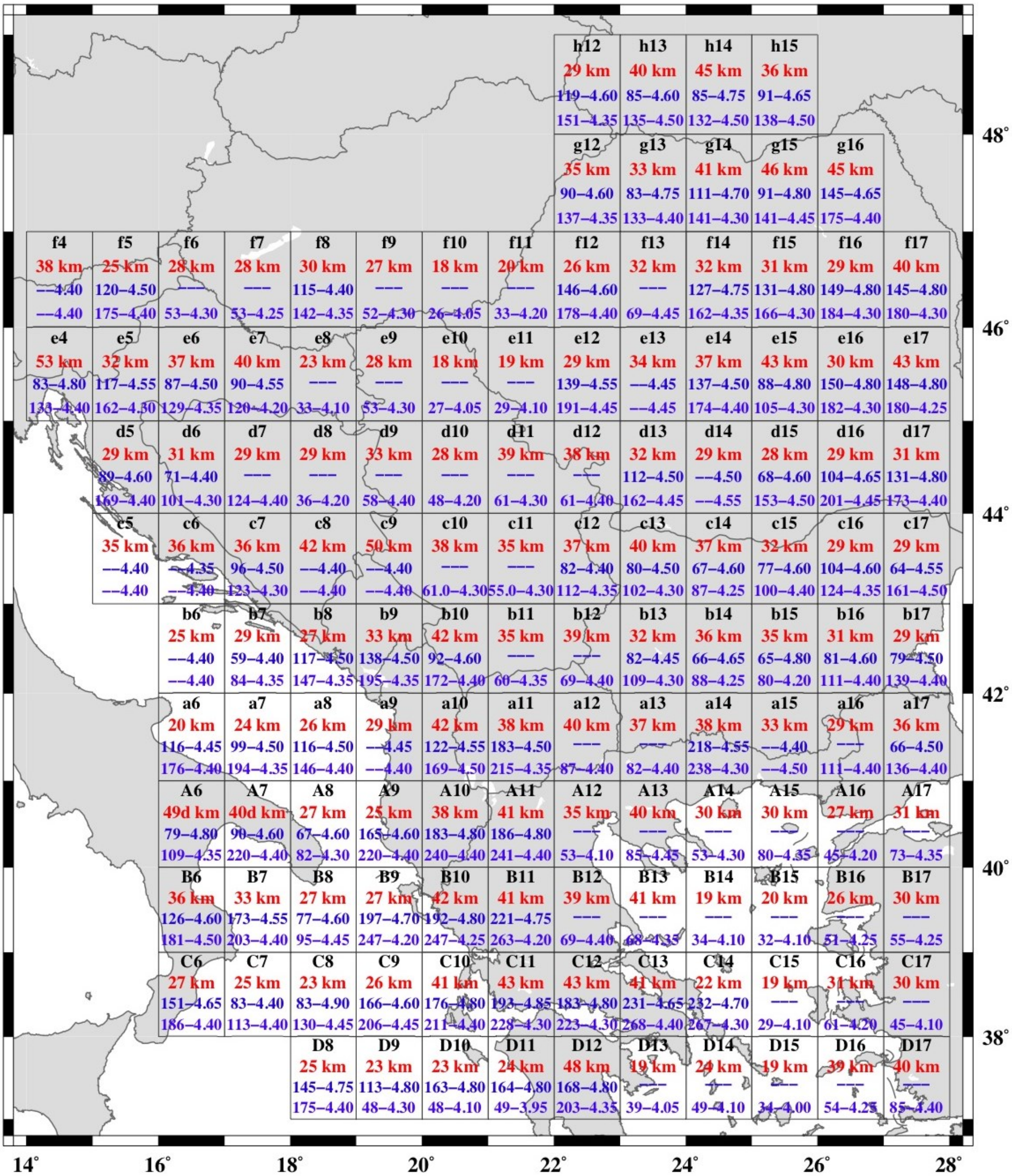

Fig. 1. Balkan Peninsula region with studied cells represented by rectangles. See the text for details.

## Results

In the Balkan Peninsula region 133 cells have been processed (Fig. 1). The representative solutions in the westmost cells (f4, e4, e5, d5, c5, c6, b6, a6, A6, B6, C6, C7, C8, and D8) were kept fixed according to Brandmayr et al. (2010) because application of different smoothing algorithms in Italian region selected the same models as the representative solution in these cells.

Independent information about the depth of Moho boundary from Grad et al. (2009) and Tesauro et al. (2010) was used for evaluation of the cellular models. Depth distribution of seismicity was used to resolve some ambiguities in layer's definition, using the revised ISC catalogue (ISC, 2007) for the period 1904–June 2005. Hypocenters were grouped in depth intervals (4 km for the crustal seismicity and 10 km for the mantle seismicity), considering the uncertainties of the hypocenter's depth calculation.

The juxtaposition of the representative one-dimensional models of each cell results in the 3D shear-wave velocity structure of the study region. Each representative cellular solution gives the preferred layer's value of $V_S$ and thickness with uncertainties, in general equal to ±1/2 of the relevant parameter step.

The results are summarized in Fig. 1. Inside every rectangle: cell's label; Moho boundary depth in km; lithosphere thickness in km and maximal $V_S$ in the lithosphere in km/s (lithosphere thickness and velocity are not defined when in the top mantle layer $V_S$ is the lowest mantle velocity); the middle depth (in km) of the mantle layer with the lowest $V_S$ in km/s. The thickness of the relevant layers in the undifferentiated mantle are substituted by "-".

## Conclusions

A structural model for the lithosphere-asthenosphere system of Balkan Peninsula region was proposed based on $V_S$-depth distribution. The multidisciplinary approach was used to better constrain the obtained model to the independent geological, geophysical and petrological information. The reconstructed picture of the lithosphere-asthenosphere system evidences the following features: the lithospheric roots in southern Carpathians and Eastern Alps, high-velocity rigid lithosphere in outer part of Eastern Carpathians, low-velocity zones under Pannonian basin with extension to the Serbian rift area, low-angle subduction under the Dinarides with high-velocity slab, low-velocity area under Aegean Sea. All these features are well consistent with the Polarized Plate Tectonics model (Doglioni and Panza, 2015).

**Резюме**

**Vs структура на земната кора и горната мантия в района на Балкански полуостров.**

*Ренета Б. Райкова[1], Джулиано Ф. Панза[2]*

Намерена е скоростната структура на литосферно-астеносферната система в района Балканския полуостров за клетки с размер 1°x1° до 350 км дълбочина. Моделите са получени, прилагайки следната поредица от методи и средства: сбор от измерени дисперсионни криви на повърхностни вълни; 2D томография на дисперсионните зависимости; нелинейна инверсия на локалните дисперсионни зависимости; изглаждащ оптимизационен метод за избор на представителен модел за всяка клетка. Тримерният скоростен модел, удовлетворяващ принципа на Окам, е получен чрез позициониране на съответните представителни клетъчни модели. Получената структура показва особеностите в редица тектонски единици от района.